\documentclass[aps,
twocolumn,
floatfix,
 amsmath,amssymb,
]{revtex4-1}

\usepackage{graphicx}
\usepackage{float}
\usepackage{bm}
\usepackage{color}

\begin{document}

\title{Practical design and performance of physical reservoir computing using hysteresis}

\author{Yuhei Yamada}
\email{yamada.y.e02c@m.isct.ac.jp}
\affiliation{%
Research Center for Autonomous Systems Materialogy (ASMat),
Institute of Integrated Research,
Institute of Science Tokyo,
4259 Nagatsuta-cho, Midori-ku, Yokohama, Kanagawa, 226-8501, Japan
}

\date{\today}

\begin{abstract}
Physical reservoir computing is an innovative idea for using physical phenomena as computational resources.
Recent research has revealed that information processing techniques can improve the performance, but for practical applications, it is equally important to study the level of performance with a simple design that is easy to construct experimentally.
We focus on a reservoir composed of independent hysteretic systems as a model suitable for the practical implementation of physical reservoir computing.
In this paper, we discuss the appropriate design of this reservoir, its performance, and its limitations.
This research will serve as a practical guideline for constructing hysteresis-based reservoirs.
\end{abstract}

\maketitle

\section{Introduction}

Reservoir computing (RC) is a method for implementing machine learning \cite{nakajima2021}.
It was introduced as a type of learning process using recurrent neural networks \cite{jaeger2001,maass2002} and has gained attention because it shows high performance despite the low learning cost.
Its distinguishing feature is that the internal layer, which is called a reservoir, is not optimized but the readout layer in data processing.
This not only reduces computational costs but also makes it possible to replace the intermediate layer with physical phenomena, known as physical reservoir computing (PRC) \cite{tanaka2019}.
PRC is expected to contribute to saving computing resources and to applications in computing devices adapted to real-world environments.
One issue with PRC is that it is difficult to achieve high performance compared with numerical RC due to physical constraints.
Recent research has shown that performance can be significantly improved by innovating information processing techniques, such as the multiplexing \cite{appeltant2011,rohm2018}.
On the other hand, when considering practical applications, there is a demand to keep information processing to a minimum.
Thus, improving performance by controlling the design of the reservoir is important as well \cite{rodan2011,andrecut2017}.
In this direction, attempts have been made to create reservoirs with good performance by composing reservoirs using multiple independent dynamics \cite{gauthier2021,wang2024}.
This method is expected to be suitable for the experimental construction of reproducible physical reservoirs compared with reservoirs that measure multiple states from a single complex system.
From the viewpoint of materials science, we focus on a reservoir using hysteresis \cite{caremel2024}.
Hysteresis is observed in a variety of materials and it is controllable \cite{krasno1989,tan2001}, making it well suited for reservoir design applications.
In this study, we propose a reservoir model comprising multiple hysteresis systems, and discuss the appropriate design, its performance, and its limitations.

\section{Model}

\begin{figure}
\begin{center}
\resizebox{0.4\textwidth}{!}{%
  \includegraphics{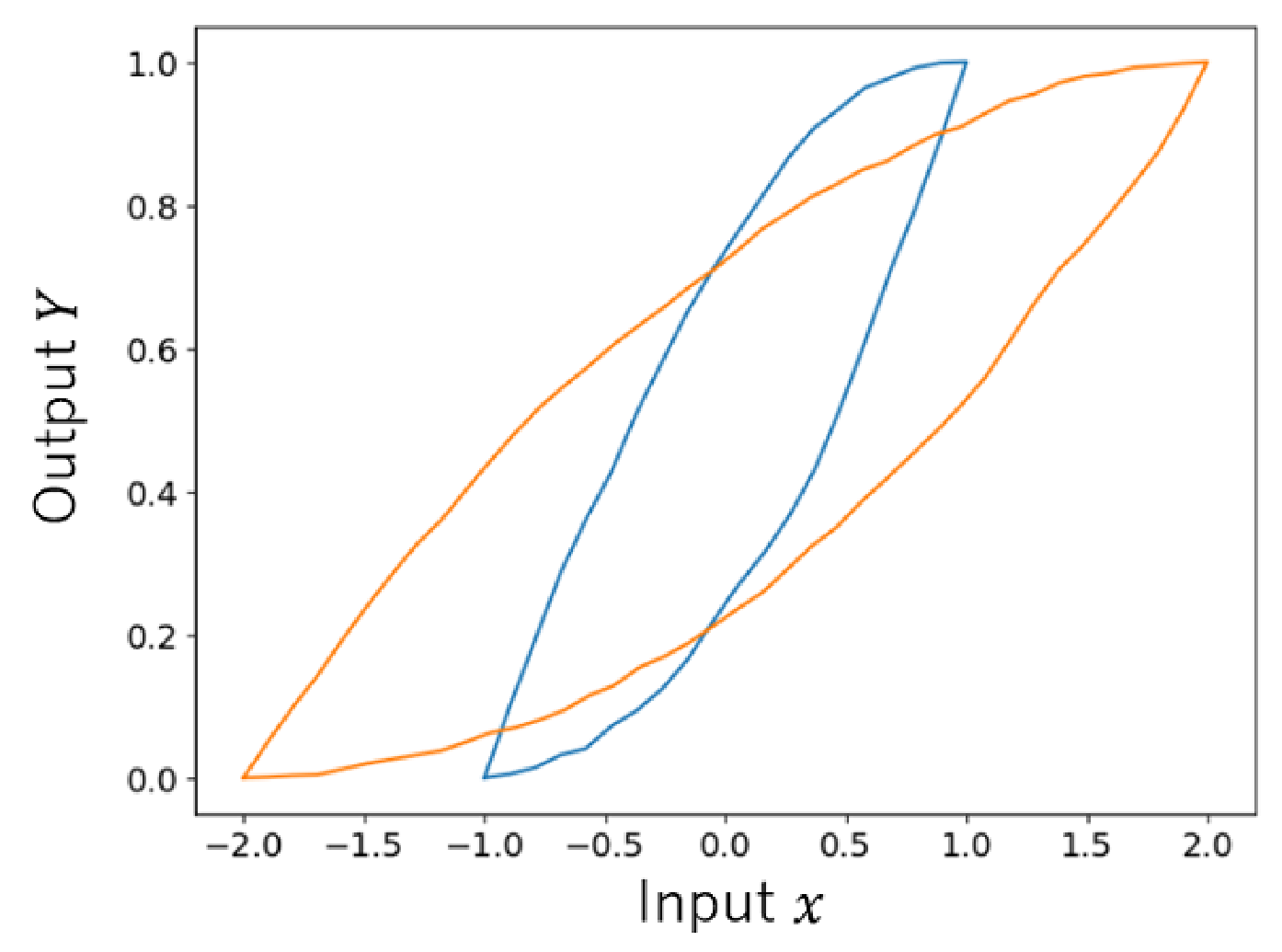}
}
\caption{
(Color online) Hysteresis examples obtained from the Preisach model.
$N_h=10^3$ and the range of $(\alpha_i,\beta_i)$ are $[-1,1]$ (blue) and $[-2,2]$ (orange), respectively.
}
\label{preisach}
\end{center}
\end{figure}

First, we briefly introduce the mathematical formulation of RC \cite{inubushi2021}.
The most basic problem of RC is imitating a target system’s output to a sequential input.
This is called an imitation task, and by setting the input and output appropriately, it can be applied to other tasks such as prediction, computation, classification etc. \cite{wringe2025}.
Here, we denote the input and output of the target system in discrete time $t$ as $u(t)$ and $y(t)$, respectively.
Assume the reservoir's state at $t$ is fully characterized by the $N$-dimensional vector, $\bm{r}(t)=\{r_i(t)\}_{i=1}^N$.
With a map, $\bm{F}$, which corresponds to the reservoir dynamics, the time evolution is described as follows:
\begin{equation}
    \bm{r}(t) = \bm{F}(\bm{r}(t-1), u(t)), \label{r_evolution}
\end{equation}
with the initial condition $\bm{r}(0)$.
Then, the output of the reservoir, $\hat{y}(t)$, is calculated by the weighted summation of $M$ picked reservoir's states as
\begin{equation}
    \hat{y}(t) = \sum_{j=1}^M w_j r_j(t), \label{read}
\end{equation}
where $\{w_j\}_{j=1}^M$ is calculated to approximate $y(t)$ by $\hat{y}(t)$.
Here, $M$ is not necessarily equal to $N$; the RC can be performed without knowing the entire system.
This operation (\ref{read}) is called readout and $\{w_j\}_{j=1}^M$ is called readout weight.
The optimization of $\bm{w}$ using training data corresponds to the "learning" in RC.
The training data comprise input sequences, target system output, and reservoir state.
In the simplest case, the optimization is performed using the least squares method: Let $\bm{y}=(y(1),...,y(L))^T$ be a sequence of the target system's output and $\bm{w}=(w_1,...,w_M)^T$ be a readout weight.
$\Phi_{tj} \equiv r_j(t)$ is referred  to as a design matrix, and Eq. (\ref{read}) is described as $\hat{\bm{y}} = \Phi \bm{w}$.
When one demands to minimize the squared error, $E(\bm{w})=||\hat{\bm{y}}-\bm{y} ||^2$, one obtains set of equation, $\partial_{w_j} E(\bm{w})=0 $.
Thus, the readout weight is calculated as follows:
\begin{equation}
  \bm{w} = (\Phi^T \Phi)^{-1} \Phi^T \bm{y}.  \label{weight}
\end{equation}
Once $\bm{w}$ is fixed from the training data, the output of the reservoir can be calculated from the reservoir's state in response to the given input by Eq. (\ref{read}).
If the learning is successful, the reservoir's output $\hat{y}(t)$ imitates the target output $y(t)$ that is not learned.
In this study, we focus on a reservoir using hysteretic behavior \cite{caremel2024}.
Hysteretic systems are generated by the Preisach model \cite{preisach1935,mayergoyz1991}.
The model is performed by summing the minimal hysteretic elements called hysterons.
The output of hysteron $i$ at time $t$ to an input $x$ is as follows:
\begin{equation}
 y_i(x)=
    \left\{
\begin{array}{lll}
0 & (x \leq \alpha_i) \\
1 & (x \geq \beta_i) \\
k & (\alpha_i < x < \beta_i) 
\end{array}
\right. ,
\end{equation}
where $\alpha_i$ and $\beta_i$ are constants, and $k$ takes the last value when $x$ is outside the interval $\alpha_i < x < \beta_i$.
With a set of random numbers $(\alpha_i,\beta_i)$, the hysteretic output is obtained as
\begin{equation}
    Y(t) = \frac{1}{N_h}\sum_{i=1}^{N_h} y_i(x(t)),
\end{equation}
where $N_h$ is the number of hysterons.
Figure \ref{preisach} shows the typical numerical simulation results.
$N_h=10^3$ is fixed throughout this paper.
In each simulation, $(\alpha_i,\beta_i)$ are chosen as pairs of random numbers from the uniform distribution in the range $[-1,1]$ (blue) and $[-2,2]$ (orange), respectively.
The input $x(t)$ is a sequential increase and decrease over each range with increment of 0.1.
As the output value depends on the input history, the system has fading memory, which is an essential feature to be the reservoir \cite{nakajima2021}.
We generated multiple hysteretic systems varying the range of $(\alpha_i,\beta_i)$, each of which becomes a reservoir state.
Here, $M=10$ hysteretic systems are prepared whose ranges of $(\alpha_i,\beta_i)$ are $[-\frac{d}{2}j,\frac{d}{2}j]$, with indexing $j=1,2,...,10$.
$d$ is a parameter that controls the hysteresis width.
Before training, all systems are initialized as $x(t\leq 0)=0$.

For the benchmark task, we first used the second-order NARMA (Nonlinear Auto-Regressive Moving Average) task, which is one of the standard benchmarks of RC \cite{atiya2000,wringe2025}.
In the task, the output sequence $y(t)$ to input $u'(t)$ is computed as follows:
\begin{equation}
    y(t) = a y(t-1) + b y(t-1) y(t-2) + c u'(t)^3 + d ,  \label{second_narma}
\end{equation}
where $a, b, c, d$ are constants, and $a=0.4, b=0.4, c=0.6, d=0.1$ in this paper, which is a commonly used parameter set.
As initial condition, $y(-1)=0, y(0)=0$ are set.
In the literature, the input $u'(t)$ is usually generated as a sequence of $L$ random numbers from a uniform distribution in the range [0,0.5].
Here, for convenience when actually performing physical RC, we adopt discrete values from the same range divided into 10, i.e., the input is a sequence of random numbers from (0, 0.05, ..., 0.5).

\section{Result and Discussion}

\begin{figure}
\begin{center}
\resizebox{0.45\textwidth}{!}{%
  \includegraphics{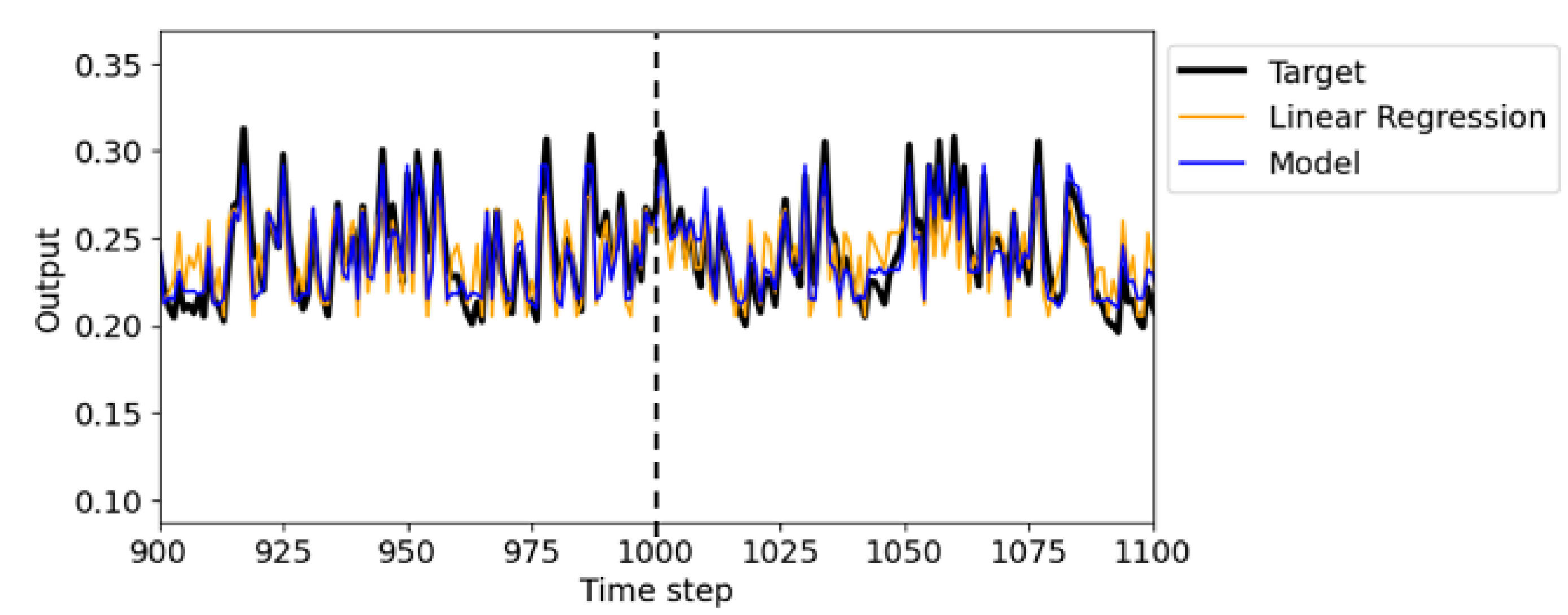}
}
\caption{
(Color online) Typical example of an imitation task for the second-order NARMA task.
The black curve indicates target output, the orange curve indicates imitation by linear regression, and the blue curve indicates imitation by the model.
The left side of the center dashed line is the training phase, and the right side of the line is the imitation phase.
}
\label{second_narma_imi}
\end{center}
\end{figure}

In the training process, the input to the reservoir is made from $u'(t)$ as $u(t)=-5+20 u'(t)$, which becomes a sequence of discrete random integers in the range of $[-5,5]$.
Depending on $d$, the input range ratio over the hysteresis range changes.
When $d \geq 10$, all hysteresises cover the input range, whereas when $d < 1$, even the largest hysteresis ($j=10$) partially covers the input range.
According to $u(t)$, each system $j$ outputs $Y_j(t)$, which defines the design matrix as $\Phi_{tj}=Y_j(t)$.
With $\Phi_{tj}$ and an output sequence of the target system, $y(t)$, we obtain the readout weight $\bm{w}$ by Eq. (\ref{weight}).
If the learning is successful, the behavior of the target system that is not learned can be predicted.
Figure \ref{second_narma_imi} shows the example of the result for the case of $d=5$.
For comparison, we used the linear regression model (LR), which assumes $\hat{y}(t)=w^1_{LR} \times u(t)+w^0_{LR}$, where $w^0_{LR}$ and $w^1_{LR}$ are regression constants.
Here, the training was performed from $t=0$ to $t=1000$, and the imitated output was calculated from $t=1001$ to $t=2000$.
To quantify the precision of imitation, we calculated the normalized mean squared error as follows:
\begin{equation}
    \mathrm{NMSE} \equiv \frac{E[ y(t)-\hat{y}(t)^2 ]}{E[ (y(t)-E[y(t)])^2 ]}.
\end{equation}
This is one of the standard metrics for evaluating imitation accuracy \cite{nakajima2021, wringe2025}.
It takes the value of 1 when the target system is approximated by the average value, and it takes a smaller value when the imitation is better.
We calculated it in the imitation sequence range, $1001 \leq t \leq 2000$.
In the case of Fig. \ref{second_narma_imi}, $\mathrm{NMSE}=0.432$ for the linear regression model and $\mathrm{NMSE}=0.201$ for the proposed model, respectively.

\begin{figure}
\resizebox{0.35\textwidth}{!}{%
  \includegraphics{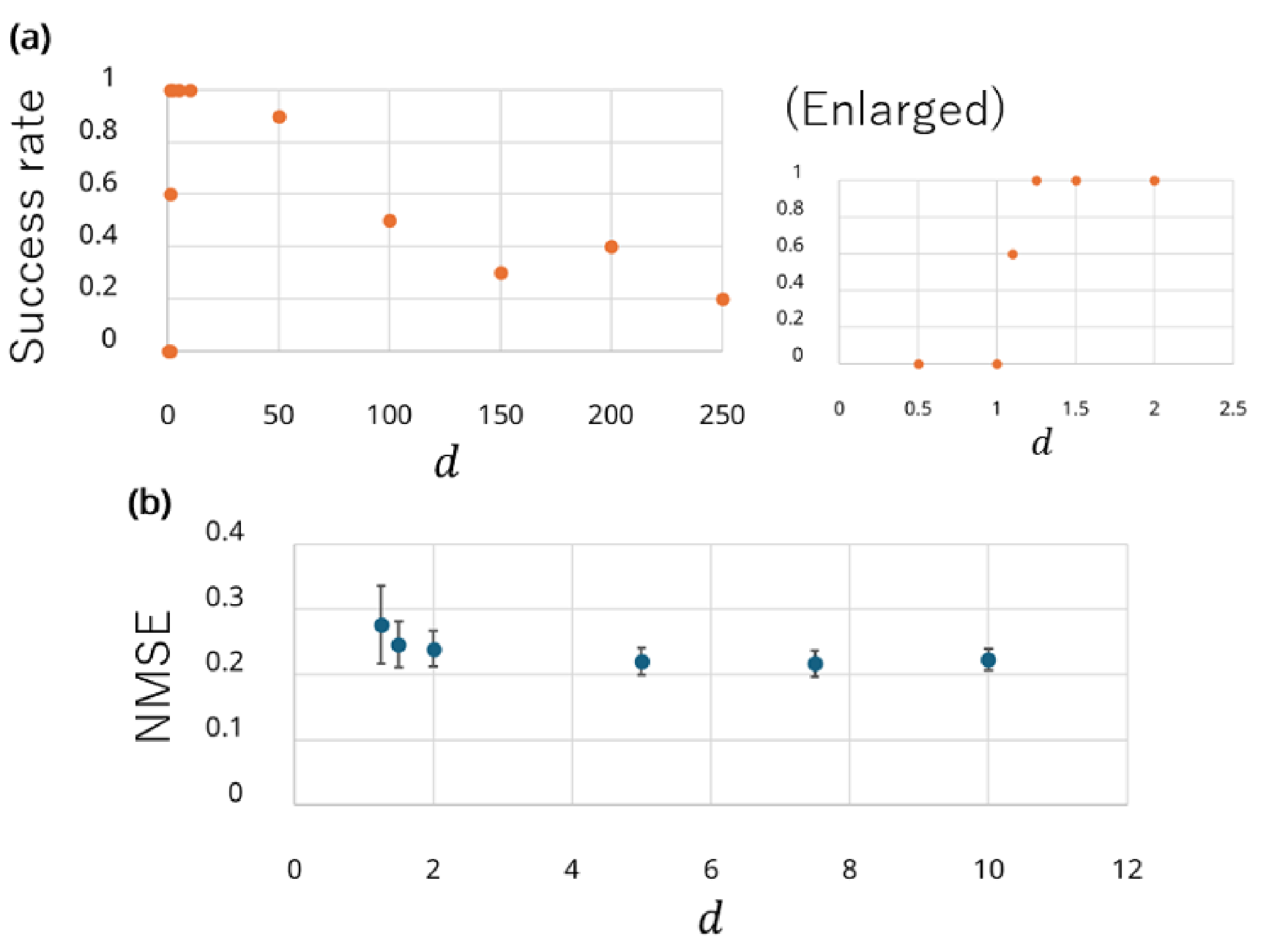}
}
\caption{
(Color online) Dependence of model performance on $d$.
Parameters other than $d$ are the same as Fig. \ref{second_narma_imi}.
(a) The percentage of trials in which the model's NMSE was lower than that of the linear regression model (LR).
(b) NMSE of the model. Average of 10 trials. Error bars indicate standard deviation.
}
\label{performance}
\end{figure}

\begin{figure}
\resizebox{0.35\textwidth}{!}{%
  \includegraphics{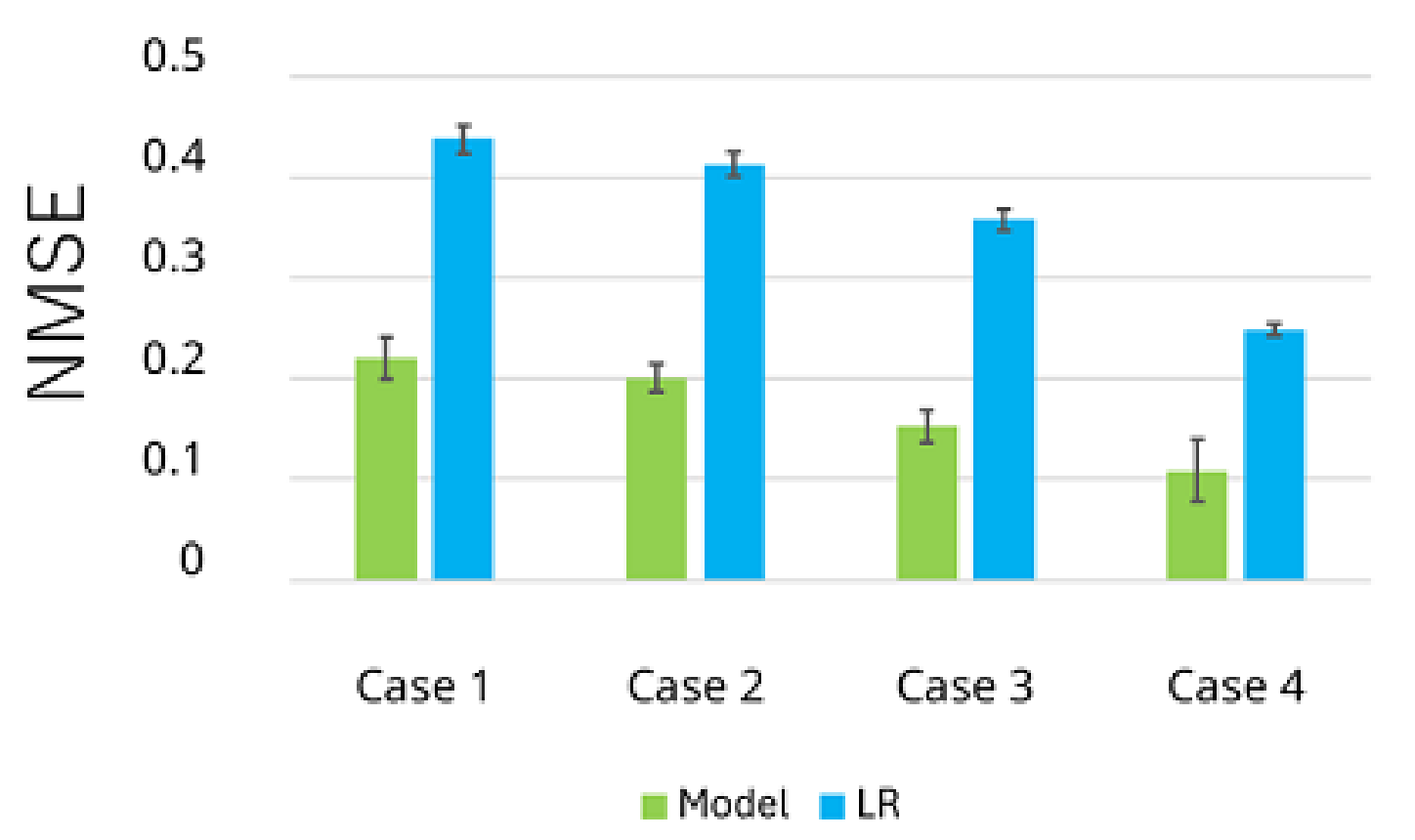}
}
\caption{
(Color online) Imitation performance of the model for the second-order NARMA task varying parameters.
Values are the average of 10 trials, and the error bars show the standard deviations.
$(a,b,c,d)=(0.4, 0.4, 0.6,0.1)$ (Case 1), $(0.4, 0.4, 0.2, 0.1)$ (Case 2), $(0.4, 0.2, 0.6, 0.1)$ (Case 3), and $(0.2, 0.4, 0.6, 0.1)$ (Case 4).
The parameters of the reservoir model are the same as Fig. \ref{second_narma_imi}.
}
\label{nmse}
\end{figure}

The imitation performance of this model depends on the value of $d$.
Figure \ref{performance} (a) shows the $d$-dependence of the percentage of times the model's NMSE was lower than LR's among 10 trials with different random numbers, which we call the success rate.
Imitation fails when $d$ is too small or too large.
In our simulation, the success rate = 1 is achieved in the range $1.25 \leq d\leq 10$.
This result suggests that both sensitive and insensitive memory effects are required for imitation of the time series.
Figure \ref{performance} (b) shows the $d$ dependence of the model’s NMSE when imitation is successful, indicating that the accuracy does not change significantly when $d$ is sufficiently greater.
Next, we examine the performance of the model by varying the target system parameters.
Figure \ref{nmse} shows comparison of the imitation performance for the four parameter sets of the second-order NARMA task for $d=5$ case.
The model is roughly twice as accurate as the LR model for second-order NARMA systems, even if the parameters are changed.

\begin{figure}
\resizebox{0.5\textwidth}{!}{%
  \includegraphics{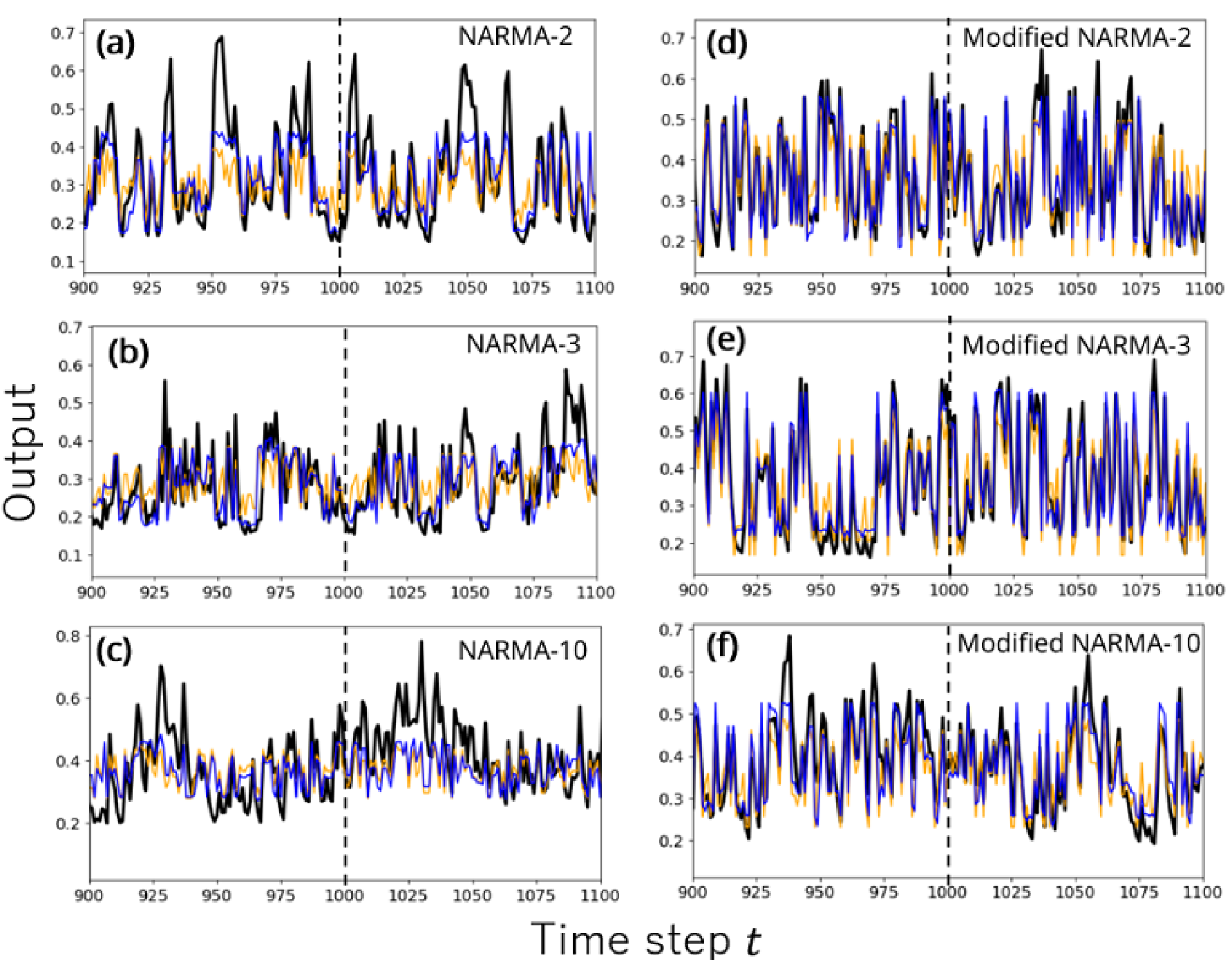}
}
\caption{
(Color online) Typical example of an imitation task for the NARMA-$N$ task.
The black curve indicates target output, the orange curve indicates imitation by linear regression, and the blue curve indicates imitation by the model.
The left side of the center dashed line is the training phase, and the right side of the line is the imitation phase.
}
\label{narma_N_imi}
\end{figure}

\begin{figure}
\resizebox{0.4\textwidth}{!}{%
  \includegraphics{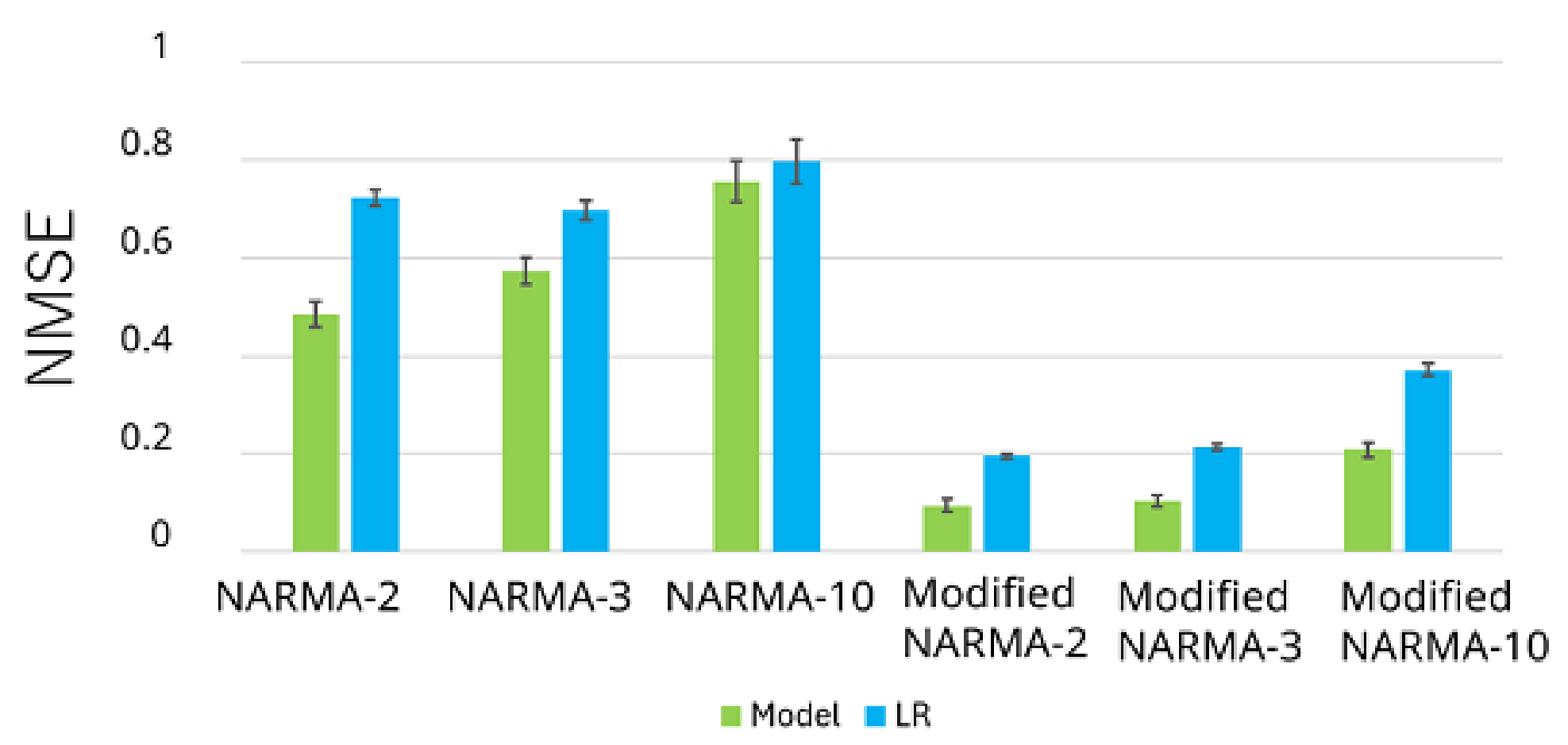}
}
\caption{
(Color online) Imitation performance of the model for the NARMA-$N$ task.
Values are the average of 10 trials, and the error bars show the standard deviations.
}
\label{nmse2}
\end{figure}

Furthermore, we investigated the applicability of this model to different benchmarks.
The NARMA-$N$ system is a well-known benchmark that is similar to, but different from, the second-order NARMA \cite{wringe2025,rodan2011}.
With a sequential input $u'(t)$, the output $y(t)$ is calculated as follows:
\begin{gather}
    y(t) = \alpha y(t-1) + \beta y(t-1) \left( \sum_{i=0}^{N-1} y(t-1-i) \right) \notag \\ 
     +\gamma u'(t-N) u'(t-1) + \delta ,  \label{narma_N}
\end{gather}
where $\alpha, \beta, \gamma, \delta$ are constants.
The initial condition is set as $y(t\leq 0)=0$ throughout this paper.
The differences between NARMA-2 and second-order NARMA are the way the moving average is taken in the second term on the right-hand side and how the input is entered in the third term. 
A notable difference is that in the NARMA-$N$ system, the input is introduced by multiplication with the value from $N-1$ steps earlier, i.e., the input has the delay effect.
As a commonly used parameter set, we set $\alpha=0.3, \beta=0.05, \gamma=1.5, \delta=0.1$, and generated the input $u'(t)$ similarly as the case of second-order NARMA.
The same $d=5$ reservoir model is used for the imitation task.
Figures \ref{narma_N_imi} (a), (b), and (c) show typical results for $N=2,3$ and 10, respectively.
Compared with the case shown in Fig. \ref{second_narma_imi}, the imitation appears less successful.
Indeed, the NMSE for the cases is greater than that for the case of second-order NARMA, as shown in Fig. \ref{nmse2}.
To elucidate the cause of the performance difference, we created a modified NARMA-$N$ system whose input is modified so that it is not dependent on the past value:
\begin{gather}
    y(t) = \alpha y(t-1) + \beta y(t-1) \left( \sum_{i=0}^{N-1} y(t-1-i) \right) \notag \\ 
     +\gamma u'^2(t-1) + \delta.  \label{mod_narma_N}
\end{gather}
We performed the imitation task using the same parameters as the standard NARMA case, and the results are shown in Figs. \ref{narma_N_imi} (d) and (e).
Only in the case of modified NARMA-10, Fig. \ref{narma_N_imi} (f), $\gamma=1.0$ is set because the output diverges with the original parameter set.
Figure \ref{nmse2} shows NMSEs for them.
Compared with the standard NARMA cases, NMSEs become significantly smaller for the modified cases.
Among the modified NARMA cases, NMSE increases with increasing $N$, but the value for our reservoir model is still approximately less than 0.2.
This shows that the model can imitate the system to some accuracy even if the target dynamics has the memory effect in the moving average as long as the input does not have the delay effect.

\section{Conclusion}

In this paper, the design and performance of a hysteretic reservoir are investigated.
The Preisach model was used to model the hysteresis systems, and the reservoir was composed of multiple systems of different hysteresis widths, each of which was used as a reservoir state.
Numerical simulations revealed that successful learning of this reservoir requires systems with various range of hysteresis over a range of inputs, and in that case, it can imitate the second-order NARMA system.
On the other hand, the imitation accuracy becomes significantly poor for the NARMA-$N$ systems.
The cause is estimated to be the delay effect of the input in NARMA-$N$ systems, and a simulation revealed that the accuracy improves when the input is modified not to have the delay effect.
This result provides insight into the practical implementation of the hysteretic reservoir.
Note that the imitation performance of this reservoir is not high.
For example, a reservoir using hysteresis and multiplexing can imitate the standard NARMA-10 task, and NMSE$\simeq 0.0063$ (NRMSE=0.0794) is reported \cite{caremel2024}.
(Although the number of optimization parameters and the amount of training data are larger than our study.)
Nevertheless, since the level of accuracy required for practical use depends on the cases, elucidating the accuracy that can be achieved with simple settings must be as important as achieving high accuracy.
In this research, the Preisach model was used as a model of the hysteretic system.
This model is known to reproduce actual hysteresis behavior well with controlling parameters \cite{tan2001}, but it is not obvious whether the results in this work can be applied to all actual systems.
Experimentally constructing the reservoir and evaluating its performance is an important future work, and we believe this work will be helpful.

\begin{acknowledgments}
I am grateful to the discussion with colleagues at ASMat, Science Tokyo, especially Prof. S. Maeda.
\end{acknowledgments}

\section*{Funding}

This work was supported by JSPS KAKENHI Grant Number JP24H00712.

\section*{Data Availability Statement}

The datasets used and/or analysed during the current study available from the corresponding author on reasonable request.

\bibliography{citation}

\end{document}